\documentclass[twocolumn,showpacs,preprintnumbers,nofootinbib,prd,
superscriptaddress,10pt]{revtex4-1}

\pdfoutput=1

\usepackage[bookmarks, pdfusetitle, pdfauthor={Stefano Schmidt}]{hyperref}
\usepackage{amsmath,amssymb}
\usepackage{amsfonts}
\usepackage{mathtools}
\usepackage[normalem]{ulem}
\usepackage{textcomp}
\usepackage{enumitem}
\usepackage{bm}
\usepackage{bbm}
\usepackage{afterpage}
\usepackage{graphicx}
\graphicspath{{img/}} 
\usepackage{tensor}

\usepackage{tabularx}
\usepackage{makecell}
\usepackage{multirow}
\usepackage{booktabs}
\usepackage{orcidlink}

\usepackage{layouts}
\usepackage[utf8]{inputenc}
\usepackage{blindtext}
\usepackage{graphicx}
\usepackage{siunitx}
\begin{document}

\begin{abstract}
	Leveraging the features of the GstLAL pipeline, we present the results of a matched filtering search for asymmetric binary black hole systems with heavily misaligned spins in LIGO and Virgo data taken during the third observing run. Our target systems show strong imprints of precession whereas current searches have non-optimal sensitivity in detecting them.
	After measuring the sensitivity improvement brought by our search over standard spin-aligned searches, we report the detection of 30 gravitational wave events already discovered in the latest version of the Gravitational Wave Transient Catalog. However, we do not find any additional significant gravitational wave candidates.
	Our results allow us to place an upper limit of $R_{90\%} = 0.28^{+0.33}_{-0.04}\;\; \mathrm{Gpc^{-3}yr^{-1}}$ on the merger rate of a hypothetical subpopulation of asymmetric, heavily precessing signals, not identified by other searches. Since our upper limit is consistent with the latest rate estimates from the LIGO-Virgo-KAGRA collaboration, our findings rule out the existence of a yet-to-be-discovered population of precessing binaries.
\end{abstract}


 \title{Searching for asymmetric and heavily precessing Binary Black Holes in the gravitational wave data from the LIGO third Observing Run}
	\author{Stefano Schmidt \orcidlink{0000-0002-8206-8089}}
	\email{s.schmidt@uu.nl}
	\affiliation{Nikhef, Science Park 105, 1098 XG, Amsterdam, The Netherlands}
	\affiliation{Institute for Gravitational and Subatomic Physics (GRASP),
Utrecht University, Princetonplein 1, 3584 CC Utrecht, The Netherlands}
        
	\author{Sarah Caudill \orcidlink{0000-0002-8927-6673}}
	\affiliation{Department of Physics, University of Massachusetts, Dartmouth, MA 02747, USA}
	\affiliation{Center for Scientific Computing and Data Science Research, University of Massachusetts, Dartmouth, MA 02747, USA}
    
	\author{Jolien D. E. Creighton \orcidlink{0000-0003-3600-2406}}
	\affiliation{Leonard E.\ Parker Center for Gravitation, Cosmology, and Astrophysics, University of Wisconsin-Milwaukee, Milwaukee, WI 53201, USA}

	\author{Leo Tsukada  \orcidlink{0000-0003-0596-5648}}
	\affiliation{Department of Physics, The Pennsylvania State University, University Park, PA 16802, USA}
	\affiliation{Institute for Gravitation and the Cosmos, The Pennsylvania State University, University Park, PA 16802, USA}

	\author{Anushka Doke}
	\affiliation{Center for Scientific Computing and Data Science Research, University of Massachusetts, Dartmouth, MA 02747, USA}
	\affiliation{International Centre for Theoretical Sciences, Tata Institute of Fundamental Research, Bengaluru 560089, India}
	
	\author{Manvi Jain}
	\affiliation{Department of Physics, University of Massachusetts, Dartmouth, MA 02747, USA}
	
	\author{Anarya Ray \orcidlink{0000-0002-7322-4748}}
	\affiliation{Leonard E.\ Parker Center for Gravitation, Cosmology, and Astrophysics, University of Wisconsin-Milwaukee, Milwaukee, WI 53201, USA}

	\author{Shomik Adhicary}
	\affiliation{Department of Physics, The Pennsylvania State University, University Park, PA 16802, USA}
	\affiliation{Institute for Gravitation and the Cosmos, The Pennsylvania State University, University Park, PA 16802, USA}

	\author{Pratyusava Baral \orcidlink{0000-0001-6308-211X}}
	\affiliation{Leonard E.\ Parker Center for Gravitation, Cosmology, and Astrophysics, University of Wisconsin-Milwaukee, Milwaukee, WI 53201, USA}

	\author{Amanda Baylor \orcidlink{0000-0003-0918-0864}}
	\affiliation{Leonard E.\ Parker Center for Gravitation, Cosmology, and Astrophysics, University of Wisconsin-Milwaukee, Milwaukee, WI 53201, USA}

	\author{Kipp Cannon \orcidlink{0000-0003-4068-6572}}
	\affiliation{RESCEU, The University of Tokyo, Tokyo, 113-0033, Japan}

	\author{Bryce Cousins \orcidlink{0000-0002-7026-1340}}
	\affiliation{Department of Physics, University of Illinois, Urbana, IL 61801 USA}
	\affiliation{Department of Physics, The Pennsylvania State University, University Park, PA 16802, USA}
	\affiliation{Institute for Gravitation and the Cosmos, The Pennsylvania State University, University Park, PA 16802, USA}

	\author{Becca Ewing}
	\affiliation{Department of Physics, The Pennsylvania State University, University Park, PA 16802, USA}
	\affiliation{Institute for Gravitation and the Cosmos, The Pennsylvania State University, University Park, PA 16802, USA}

	\author{Heather Fong}
	\affiliation{Department of Physics and Astronomy, University of British Columbia, Vancouver, BC, V6T 1Z4, Canada}
	\affiliation{RESCEU, The University of Tokyo, Tokyo, 113-0033, Japan}
	\affiliation{Graduate School of Science, The University of Tokyo, Tokyo 113-0033, Japan}

	\author{Richard N. George \orcidlink{0000-0002-7797-7683}}
	\affiliation{Center for Gravitational Physics, University of Texas at Austin, Austin, TX 78712, USA}

	\author{Patrick Godwin}
	\affiliation{LIGO Laboratory, California Institute of Technology, MS 100-36, Pasadena, California 91125, USA}
	\affiliation{Department of Physics, The Pennsylvania State University, University Park, PA 16802, USA}
	\affiliation{Institute for Gravitation and the Cosmos, The Pennsylvania State University, University Park, PA 16802, USA}

	\author{Chad Hanna}
	\affiliation{Department of Physics, The Pennsylvania State University, University Park, PA 16802, USA}
	\affiliation{Institute for Gravitation and the Cosmos, The Pennsylvania State University, University Park, PA 16802, USA}
	\affiliation{Department of Astronomy and Astrophysics, The Pennsylvania State University, University Park, PA 16802, USA}
	\affiliation{Institute for Computational and Data Sciences, The Pennsylvania State University, University Park, PA 16802, USA}

	\author{Reiko Harada}
	\affiliation{RESCEU, The University of Tokyo, Tokyo, 113-0033, Japan}
	\affiliation{Graduate School of Science, The University of Tokyo, Tokyo 113-0033, Japan}

	\author{Yun-Jing Huang \orcidlink{0000-0002-2952-8429}}
	\affiliation{Department of Physics, The Pennsylvania State University, University Park, PA 16802, USA}
	\affiliation{Institute for Gravitation and the Cosmos, The Pennsylvania State University, University Park, PA 16802, USA}

	\author{Rachael Huxford}
	\affiliation{Department of Physics, The Pennsylvania State University, University Park, PA 16802, USA}
	\affiliation{Institute for Gravitation and the Cosmos, The Pennsylvania State University, University Park, PA 16802, USA}

	\author{Prathamesh Joshi \orcidlink{0000-0002-4148-4932}}
	\affiliation{Department of Physics, The Pennsylvania State University, University Park, PA 16802, USA}
	\affiliation{Institute for Gravitation and the Cosmos, The Pennsylvania State University, University Park, PA 16802, USA}

	\author{James Kennington \orcidlink{0000-0002-6899-3833}}
	\affiliation{Department of Physics, The Pennsylvania State University, University Park, PA 16802, USA}
	\affiliation{Institute for Gravitation and the Cosmos, The Pennsylvania State University, University Park, PA 16802, USA}

	\author{Soichiro Kuwahara}
	\affiliation{RESCEU, The University of Tokyo, Tokyo, 113-0033, Japan}
	\affiliation{Graduate School of Science, The University of Tokyo, Tokyo 113-0033, Japan}

	\author{Alvin K. Y. Li \orcidlink{0000-0001-6728-6523}}
	\affiliation{LIGO Laboratory, California Institute of Technology, Pasadena, CA 91125, USA}

	\author{Ryan Magee \orcidlink{0000-0001-9769-531X}}
	\affiliation{LIGO Laboratory, California Institute of Technology, Pasadena, CA 91125, USA}

	\author{Duncan Meacher \orcidlink{0000-0001-5882-0368}}
	\affiliation{Leonard E.\ Parker Center for Gravitation, Cosmology, and Astrophysics, University of Wisconsin-Milwaukee, Milwaukee, WI 53201, USA}

	\author{Cody Messick}
	\affiliation{MIT Kavli Institute for Astrophysics and Space Research, Massachusetts Institute of Technology, Cambridge, MA 02139, USA}

	\author{Soichiro Morisaki \orcidlink{0000-0002-8445-6747}}
	\affiliation{Institute for Cosmic Ray Research, The University of Tokyo, 5-1-5 Kashiwanoha, Kashiwa, Chiba 277-8582, Japan}
	\affiliation{Leonard E.\ Parker Center for Gravitation, Cosmology, and Astrophysics, University of Wisconsin-Milwaukee, Milwaukee, WI 53201, USA}

	\author{Debnandini Mukherjee  \orcidlink{0000-0001-7335-9418}}
	\affiliation{NASA Marshall Space Flight Center, Huntsville, AL 35811, USA}
	\affiliation{Center for Space Plasma and Aeronomic Research, University of Alabama in Huntsville, Huntsville, AL 35899, USA}

	\author{Wanting Niu \orcidlink{0000-0003-1470-532X}}
	\affiliation{Department of Physics, The Pennsylvania State University, University Park, PA 16802, USA}
	\affiliation{Institute for Gravitation and the Cosmos, The Pennsylvania State University, University Park, PA 16802, USA}

	\author{Alex Pace}
	\affiliation{Department of Physics, The Pennsylvania State University, University Park, PA 16802, USA}
	\affiliation{Institute for Gravitation and the Cosmos, The Pennsylvania State University, University Park, PA 16802, USA}

	\author{Cort Posnansky}
	\affiliation{Department of Physics, The Pennsylvania State University, University Park, PA 16802, USA}
	\affiliation{Institute for Gravitation and the Cosmos, The Pennsylvania State University, University Park, PA 16802, USA}

	\author{Surabhi Sachdev \orcidlink{0000-0002-0525-2317}}
	\affiliation{School of Physics, Georgia Institute of Technology, Atlanta, GW 30332, USA}
	\affiliation{Leonard E.\ Parker Center for Gravitation, Cosmology, and Astrophysics, University of Wisconsin-Milwaukee, Milwaukee, WI 53201, USA}

	\author{Shio Sakon \orcidlink{0000-0002-5861-3024}}
	\affiliation{Department of Physics, The Pennsylvania State University, University Park, PA 16802, USA}
	\affiliation{Institute for Gravitation and the Cosmos, The Pennsylvania State University, University Park, PA 16802, USA}

	\author{Divya Singh \orcidlink{0000-0001-9675-4584}}
	\affiliation{Department of Physics, The Pennsylvania State University, University Park, PA 16802, USA}
	\affiliation{Institute for Gravitation and the Cosmos, The Pennsylvania State University, University Park, PA 16802, USA}

	\author{Urja Shah \orcidlink{0000-0001-8249-7425}}
	\affiliation{School of Physics, Georgia Institute of Technology, Atlanta, GW 30332, USA}

	\author{Ron Tapia}
	\affiliation{Department of Physics, The Pennsylvania State University, University Park, PA 16802, USA}
	\affiliation{Institute for Computational and Data Sciences, The Pennsylvania State University, University Park, PA 16802, USA}

	\author{Takuya Tsutsui \orcidlink{0000-0002-2909-0471}}
	\affiliation{RESCEU, The University of Tokyo, Tokyo, 113-0033, Japan}

	\author{Koh Ueno \orcidlink{0000-0003-3227-6055}}
	\affiliation{RESCEU, The University of Tokyo, Tokyo, 113-0033, Japan}

	\author{Aaron Viets \orcidlink{0000-0002-4241-1428}}
	\affiliation{Concordia University Wisconsin, Mequon, WI 53097, USA}

	\author{Leslie Wade}
	\affiliation{Department of Physics, Hayes Hall, Kenyon College, Gambier, Ohio 43022, USA}

	\author{Madeline Wade \orcidlink{0000-0002-5703-4469}}
	\affiliation{Department of Physics, Hayes Hall, Kenyon College, Gambier, Ohio 43022, USA}

	\maketitle
	


\section{Introduction}

When it comes to binary black holes (BBHs) systems, the orientation of their spin vectors plays a crucial role in the dynamics of their inspiral and the resulting gravitational waves (GWs) they emit. In particular, when the spins of the two black holes are misaligned with the orbital angular momentum, the binary system exhibits a phenomenon known as precession~\cite{Apostolatos:1994mx}. This means that the orbital plane itself wobbles and changes orientation as the two black holes spiral towards each other. Precession introduces unique signatures in the gravitational wave signals, making them not only more complex but also scientifically interesting. The effect of precession is larger in systems with large spin magnitudes or significant spin misalignment as well as in those with a substantial mass asymmetry between the two black holes and in systems observed with an edge-on inclination.

The observation of precession in BBHs can have a large scientific impact.
First of all, the degree of spin misalignment can provide a better understanding of the BBH formation mechanisms~\cite{Farr:2017gtv, Farr:2017uvj, Johnson-McDaniel:2021rvv, Vitale:2022dpa}, allowing to tune the details of current models which predict either an isolated binary evolution or a dynamic interactions in dense stellar environments. Furthermore, precession affects the waveform's morphology, potentially allowing us to better measure key parameters like the luminosity distance of the source, which in turn can yield to more precise constraints of the Hubble constant~\cite{Vitale:2018wlg, Yun:2023ygz} through GW observations only.

The detection of the more than $90$ BBHs events~\cite{LIGOScientific:2018mvr, LIGOScientific:2020ibl, LIGOScientific:2021usb, KAGRA:2021vkt} of the GW Transient Catalog (GWTC) by the LIGO~\cite{LIGOScientific:2014pky}, Virgo~\cite{VIRGO:2014yos} and KAGRA~\cite{KAGRA:2020tym} (LVK) collaboration has allowed to observe precession as a statistical property of the BBH population~\cite{LIGOScientific:2020kqk, KAGRA:2021duu}. At the same time, very few signals show conclusive evidence for largely misaligned spins~\cite{Hoy:2021dqg}, interesting exceptions including GW190521~\cite{LIGOScientific:2020iuh}, GW191109~\cite{Zhang:2023fpp} and GW200129~\cite{Hannam:2021pit}.

So many detections were made possible thanks to sophisticated matched-filtering pipelines~\cite{Allen:2005fk, Privitera:2013xza, DalCanton:2014hxh, Usman:2015kfa, Nitz:2017svb, Davies:2020tsx, Adams:2015ulm, Aubin:2020goo, Luan:2011qx, Chu:2017ovg, Chu:2020pjv, Capano:2016dsf, Venumadhav:2019tad}, which correlate a large number of signal templates with the interferometric data, looking for a potential match.
However, all such pipelines are designed to search for {\it aligned-spin} systems, thus completely neglecting precession. While this choice ensures that the search can run at a moderate computational cost, aligned-spin pipelines have a poor sensitivity for systems presenting a very large amount of precession, such as very asymmetric BBHs with misaligned spins~\cite{DalCanton:2014qjd, Harry:2016ijz}.

The poor sensitivity of traditional matched-filter pipelines towards precessing systems poses an important question, which we address in this {\it Letter}. 
Is the observed lack of strongly precessing signals due to their rarity? Or, rather, is it caused by the limited sensitivity of current searches to such extreme signals? To discern between the two alternatives, it is important to explore thoroughly the precessing BBH parameter space, using the most sensitive possible matched filtering search pipeline.
In this spirit, in~\cite{Schmidt:2024jbp} we have developed a new search method, specifically tailored to detect precessing BBH signals, based on the GstLAL search pipeline~\cite{Messick:2016aqy, Sachdev:2019vvd, Hanna:2019ezx, cannon2020gstlal, Ewing:2023qqe, Tsukada:2023edh} and we identified a region of the BBH parameter space where a sensitivity improvement of up to $120\%$ is within reach. See also~\cite{McIsaac:2023ijd} for a promising alternative.
Here, we leverage our results~\cite{Schmidt:2024jbp} to search the LIGO-Virgo publicly available data from the third observing run O3~\cite{KAGRA:2023pio} for precessing BBH signals with component masses ${m_1 \in [15, 70]\SI{}{M_\odot}}$ and ${m_2 \in [3, 10]\SI{}{M_\odot}}$, with a mass ratio restricted to ${q = m_1/m_2 \in [5, 12]}$.

After briefly reviewing our method in Sec.~\ref{sec:methods}, we present the search results in Sec.~\ref{sec:results}. While we do not detect any novel signal, our results allow us to place an upper limit on the astrophysical merger rate of a hypothetical subpopulation of asymmetric, heavily precessing signals, not detected by past searches. As we will argue below, our findings suggest that the observed lack of heavily precessing signals is indeed due to their rarity and not to the poor sensitivity of the search pipelines.

\section{Methods}\label{sec:methods}

At its core, the GstLAL pipeline, as well as any other matched-filtering pipeline~\cite{Allen:2005fk}, employs the technique of matched filtering to correlate the interferometer data with a template signal. The procedure is repeated for many templates, which are stored in a {\it template bank}~\cite{Sathyaprakash:1991mt, Dhurandhar:1992mw, Owen:1998dk, Babak:2006ty, Cokelaer:2007mv}, designed to adequately cover the space of physical signals of interest.
A potential GW candidate is characterized by a large correlation between the interferometer data and a template and, for each potential candidate, the GstLAL pipeline records a {\it trigger}, notably consisting of signal-to-noise ratio (SNR) and signal-consistency test value $\xi^2$~\cite{Messick:2016aqy}. The SNR measures the ``loudness'' of the signal as compared to the detector's noise floor, whereas the $\xi^2$ value quantifies the discrepancy between the measured and expected time dependent correlation of a template with some data containing a GW signal.

Each trigger, or coincidence of triggers happening simultaneously across multiple detectors, is ranked according to the likelihood ratio $\mathcal{L}$~\cite{Cannon:2015gha, Hanna:2019ezx, Tsukada:2023edh}. This ratio represents the probability of a trigger originating from an astrophysical source compared to the probability of it being caused by detector noise.
Based on its likelihood, each trigger is assigned a False Alarm Rate (FAR), which amounts to the rate at which a trigger with a given $\mathcal{L}$ occurs in a search where no astrophysical signals are present.
Finally, each trigger is assigned a probability, $p_\text{astro}$~\cite{PhysRevD.91.023005, Kapadia:2019uut}, that it originates from an astrophysical source, using the Poisson mixture model formalism.
Although $p_\text{astro}$ has a one-to-one mapping with the FAR, it offers a more physically interpretable measure. For this reason, a detection is only claimed for candidates with $p_\text{astro} > 0.5$\cite{KAGRA:2021vkt}, which are then labeled as ``events".

To search for precessing signals, we employ the search technique introduced and implemented in~\cite{Schmidt:2024jbp}. The method enhances the matched-filtering routines described above and offers up to 120\% sensitive volume increase for very asymmetric highly precessing BBH systems, as compared to its aligned-spin counterpart.
The sensitivity improvement is made possible thanks to three main features:
\begin{itemize}
	\item {\it New metric template placement algorithm} \cite{Schmidt:2023gzj}: by reducing the bank generation time by orders of magnitude, it allows for a fast identification of a region of the parameter space where a sensitivity improvement can be achieved using feasible number of templates.
	\item {\it Updated signal consistency test} \cite{Schmidt:2024kxy}: it improves the robustness of the signal identification stage.
	\item {\it Implementation of a suitable search statistics} \cite{Harry:2017weg}: it allows for optimal SNR recovery of the candidates.
\end{itemize}
We encourage the interested reader to go through the aforementioned references for more details.

\begin{table}[t]
	
	\begin{tabular}{l|ccc}
		\toprule    
		Event name & SNR   & $\text{SNR}_\text{GWTC}$  & $\textrm{FAR} (1/\textrm{yr})$ \\
		\hline
		\midrule
		GW190408\_181802$\;\;$ & 14.29 & 14.6 & $<10^{-5}$ \\
		GW190412\_053044$\;\;$ & 17.82 & 19.8 & $<10^{-5}$ \\
		GW190512\_180714$\;\;$ & 11.88 & 12.7 & $<10^{-5}$ \\
		GW190513\_205428$\;\;$ & 11.8 & 12.5 & 0.0008 \\
		GW190519\_153544$\;\;$ & 12.69 & 15.9 & 0.0001 \\
		GW190521\_074359$\;\;$ & 22.74 & 25.9 & $<10^{-5}$ \\
		GW190527\_092055$\;\;$ & 8.79 & 8.0 & 0.1547 \\
		GW190706\_222641$\;\;$ & 12.54 & 13.4 & 0.0221 \\
		GW190707\_093326$\;\;$ & 12.68 & 13.1 & $<10^{-5}$ \\
		GW190708\_232457$\;\;$ & 12.13 & 13.4 & $<10^{-5}$ \\
		GW190720\_000836$\;\;$ & 10.55 & 10.9 & 0.0012 \\
		GW190727\_060333$\;\;$ & 11.3 & 11.7 & $<10^{-5}$ \\
		GW190728\_064510$\;\;$ & 11.92 & 13.1 & $<10^{-5}$ \\
		GW190814\_211039$\;\;$ & 20.52 & 25.3 & $<10^{-5}$ \\
		GW190828\_063405$\;\;$ & 15.93 & 16.5 & $<10^{-5}$ \\
		GW190828\_065509$\;\;$ & 10.56 & 10.2 & 0.001 \\
		GW190915\_235702$\;\;$ & 12.09 & 13.1 & 0.0015 \\
		GW190924\_021846$\;\;$ & 12.87 & 12.0 & $<10^{-5}$ \\
		GW191109\_010717$\;\;$ & 14.19 & 17.3 & $<10^{-5}$ \\
		GW191129\_134029$\;\;$ & 13.14 & 13.1 & $<10^{-5}$ \\
		GW191204\_171526$\;\;$ & 16.78 & 17.4 & $<10^{-5}$ \\
		GW191215\_223052$\;\;$ & 10.36 & 11.2 & 0.0008 \\
		GW191216\_213338$\;\;$ & 17.52 & 18.6 & $<10^{-5}$ \\
		GW191222\_033537$\;\;$ & 10.91 & 12.5 & 0.0009 \\
		GW200128\_022011$\;\;$ & 9.0 & 10.6 & 0.0059 \\
		GW200129\_065458$\;\;$ & 25.8 & 26.8 & $<10^{-5}$ \\
		GW200224\_222234$\;\;$ & 17.85 & 20.0 & $<10^{-5}$ \\
		GW200225\_060421$\;\;$ & 12.1 & 12.5 & 0.0664 \\
		GW200311\_115853$\;\;$ & 15.69 & 17.8 & $<10^{-5}$ \\
		GW200316\_215756$\;\;$ & 9.96 & 10.3 & 0.0003 \\
		\bottomrule
	\end{tabular}

	\caption{Confident events detected by our search with ${p_\textrm{astro}>0.5}$.
		For each event, we report the event name, together with the SNR and FAR. For reference, we also report the $\text{SNR}_\text{GWTC}$ reported in the Transient Catalog.}
	\label{tab:gwtc_events}
\end{table}

\begin{table}[t]
	
	\begin{tabular}{l|ccc}
		\toprule
		GPS time & SNR  & $\textrm{FAR} (1/\textrm{yr})$ &   $p_\textrm{astro}$ \\
		\hline
		\midrule
		$1252415231\;\;$ & 8.84 & 2.367 & 0.06 \\
		$1252465013\;\;$ & 8.35 & 3.2005 & 0.05 \\
		$1253504581\;\;$ & 9.02 & 4.5486 & 0.03 \\
		$1267433277\;\;$ & 7.75 & 5.4287 & 0.04 \\
		\bottomrule
	\end{tabular}
	
	\caption{Candidates not reported in GWTC, detected by our search with ${\textrm{FAR} < 6/\textrm{yr} = 1/2\, \textrm{months}}$. For each candidate, we report the SNR, FAR and $p_\textrm{astro}$. }
	\label{tab:significant_events}
\end{table}

\begin{figure}[t]
	\centering
	\includegraphics[scale = 1.]{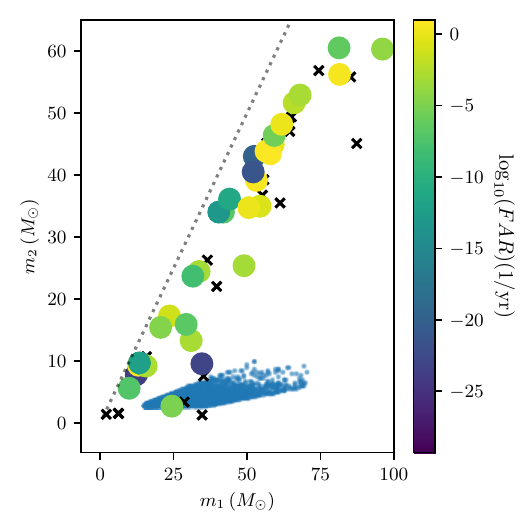}
	\caption{Median masses of the GWTC candidates as computed with parameter estimation. The GWTC candidates detected by our search are denoted by large circles, colored according to their FAR. A black cross marks the GWTC events that our search missed. The blue dots represent the templates used in our search bank.}
	\label{fig:missed_found_gwtc}
\end{figure}

\begin{figure}[t]
	\centering
	\includegraphics[scale = 1.]{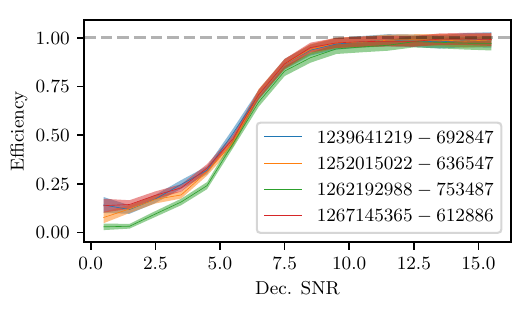}
	\caption{Efficiency of the search as a function of the decisive SNR, i.e. the maximum injected SNR between the two detectors. The efficiency is evaluated at a FAR threshold of $1/\textrm{yr}$ and it is measured on four different chunks with GPS start time and length reported on the caption separated by a dash.}
	\label{fig:efficiency}
\end{figure}

\subsection{Search setup}\label{sec:search_setup}

Our search employs the ``High $q$"\footnote{The bank was generated using a $0.9$ minimal match requirement. In the jargon of matched filtering search, this is the minimal tolerable SNR loss caused by the discreteness of the template bank. The choice of using a $0.9$ minimal match, instead of the standard $0.97$, allows us to obtain a bank with a manageable size. As discussed in~\cite{Schmidt:2024jbp}, it does not seem to have a large negative impact on the search.}
precessing template bank introduced in~\cite{Schmidt:2024jbp} with 2.3 million templates, targeting systems with component masses ${m_1 \in [15, 70]\SI{}{M_\odot}}$ and ${m_2 \in [3, 10]\SI{}{M_\odot}}$ and restricted to a mass ratio ${q = m_1/m_2 \in [5, 12]}$.

The spin vectors $\mathbf{s}_1, \mathbf{s}_2$ of the heavier and lighter black holes respectively are typically expressed in a Cartesian coordinate system where the orbital plane is orthogonal to the z-axis and the two BHs are aligned on the x-axis at a specified reference orbital frequency $f_\text{ref}$, which we set at $\SI{15}{Hz}$.
Although, in general, the effect of precession depends on the four in-plane components of the spin vectors—namely $s_\text{1x}, s_\text{1y}, s_\text{2x}$, and $s_\text{2y}$—it turns out that the GW emission of a precessing system can be accurately described using a suitable configuration in which $s_\text{1x}$ is the {\it only} non-zero in-plane spin component~\cite{Schmidt:2012rh, Schmidt:2014iyl}.
Consequently, we only consider templates where $s_\text{1y} = s_\text{2x} = s_\text{2y} = 0$, allowing only $s_\text{1x}, s_\text{1z},$ and $s_\text{2z}$ to be non-zero\footnote{Physically, this corresponds to a scenario where, at the reference frequency, only the spin of the first object is misaligned with the orbital plane, while the spin of the second object remains aligned with it.}. This approach is found to sufficiently cover the precessing parameter space, while minimizing the number of templates in the bank and reducing the computational cost of the search~\cite{Schmidt:2023gzj, Schmidt:2024jbp}.

To focus only on systems with a high precession content, we restrict our attention to templates where the magnitude of the spin vector of the first object $s_1$ falls within the range $[0.5, 0.9]$.
For the lighter object, only the $s_\text{2z}$ component of the spin is non-zero, spanning the range $[-0.99,0.99]$.
Finally, templates are chosen to have an inclination angle $\iota \in [0, \pi]$.

We filter the data only considering frequencies within the range $f \in [15,1024]\SI{}{Hz}$ with the waveform approximant \texttt{IMRPhenomXP}~\cite{Pratten:2020ceb}. Note that for our analysis we only consider {\it simple} precession, thus neglecting the effect of {\it transitional} precession~\cite{Apostolatos:1994mx, Kidder:1995zr}.
To compare the performance of our precessing search we use an aligned-spin template bank, also introduced in~\cite{Schmidt:2024jbp}, covering the same mass range and with $s_\text{1z}, s_\text{2z}\in [-0.99,0.99]$, employing $2.7 \times 10^4$ templates.

To estimate the search sensitivity and to place a limit on the merger rate of precessing signals (see Sec.~\ref{sec:rates}), we sample a number of injections using the BBH population results inferred using the latest version of the Catalog~\cite{KAGRA:2021duu}. For this purpose, we use the best fit of the Binned Gaussian Process (BGP) population model \cite{Ray:2023upk}, which allows us to easily sample BBH signals in the mass range of consideration. We sample the two spins isotropically with the magnitude of each spin constrained between $0.5$ and $0.9$. The sources are uniformly distributed in the volume between $[30, 300] \SI{}{Mpc}$ with uniformly sampled orientation, and generated with the \texttt{IMRPhenomXP} approximant.

For our search we use publicly available GW data \cite{KAGRA:2023pio} collected by the two LIGO observatories \cite{LIGOScientific:2014pky, buikema2020sensitivity, Tse:2019wcy, LIGO:2021ppb} during the third observing run (O3). Data collection periods were divided in two halves: the first half of O3 (O3a) spanned between 1 April 2019, 1500 UTC and 1 October 2019, 1500 UTC, while the second half (O3b) happened between 1 November 2019, 15:00 UTC and 27 March 2020, 17:00 UTC.
Forced by limited computational resources, we did not use Virgo data taken in the same period, due to the smaller contribution to the sensitive volume given by Virgo.

\section{Search results}\label{sec:results}

Our search detected 30 events with ${p_\text{astro}>0.5}$: they were all previously reported in the past editions of the GWTC, namely GWTC-2 \cite{LIGOScientific:2020ibl, LIGOScientific:2021usb} and GWTC-3 \cite{KAGRA:2021vkt}, released by the LIGO-Virgo-KAGRA collaboration. The confident detections are reported in Tab.~\ref{tab:gwtc_events}.
In Fig.~\ref{fig:missed_found_gwtc}, we summarize the information about the GWTC signals detected by our search.
Each event not discovered by our search is represented by a cross, while a large dot, colored by the detection $\text{FAR}$, represents a detection. Note that we report candidates found by our search up to a $\text{FAR} = 1/\text{day}$, hence some of them do not satisfy our detection criterion of ${p_\text{astro}>0.5}$ and consequently are not labeled as confident events.

As shown in Fig.~\ref{fig:missed_found_gwtc} and as further confirmed by parameter estimation studies \cite{LIGOScientific:2020ibl, KAGRA:2021vkt}, the GWTC events detected by our search lie mostly outside the region covered by our template bank. We observe that these events typically have a higher total mass than our templates, meaning they are shorter in duration.
Consequently, we conclude that the sensitivity of our search extends to higher masses beyond our intended target. This is likely due to the large number of relatively long-duration templates used, which can easily match short-duration events, albeit with sub-optimal SNR and $\xi^2$ values.
The fact that the detected events are outside the target region also explains the observation in Tab.~\ref{tab:gwtc_events} that our search measures a systematically lower SNR than the GWTC searches.

To confirm the quality of our search, we note that it recovered with high confidence the three individual events previously mentioned (GW190521, GW191109, and GW200129), for which spin misalignment measurements had been claimed by other works~\cite{LIGOScientific:2020iuh, Hannam:2021pit, Zhang:2023fpp}.

In Fig.~\ref{fig:missed_found_gwtc} we note that two events, GW191113\_071753 and GW200210\_092254, were missed by our search, even if they lie close to our target region. The first one GW191113\_071753 was not recorded by the GstLAL pipeline even in the offline broadband search\footnote{The event was found instead by the PyCBC~\cite{DalCanton:2014hxh, Usman:2015kfa, Nitz:2017svb} and Multi-Band Template Analysis (MBTA)~\cite{Adams:2015ulm, Aubin:2020goo} pipelines, although with a low significance. Similar situations, where only a few pipelines find a given event, are very common for marginally significant events.} and for this reason it is not surprising that also our precessing search failed to detect it.
On the other hand, the issue of why GW200210\_092254 was missed requires more investigation. For the moment we can postulate that this is due to a combination of the signal being at the very edge of our template bank and of the low mass of the lighter object, resulting in a signal duration longer than most of the templates in the bank. 
Unlike the other short events detected outside the bank, the longer duration of this event makes it more challenging for a template to recover the entire SNR of the system.

Besides the known GW events, our search did not detect any novel GW candidate. In Tab.~\ref{tab:significant_events}, we report a list of the candidates observed with ${\text{FAR}<2/\text{year}}$, not included in a previous Transient Catalog.

In Fig.~\ref{fig:efficiency}, we report the efficiency $\epsilon(\text{SNR}, \text{FAR})$ of our search for different bins of {\it decisive} SNR, defined as the maximum injected SNR between the two LIGO detectors. The efficiency amounts to the fraction of signals detected by the search with a given FAR threshold and it is computed using the BGP injection set defined above.
To reduce our computational footprint, we perform injections only for two chunks of data in O3a and O3b respectively, and we report the results separately for each chunk. 
The efficiencies obtained are similar to those reported in \cite{Schmidt:2024jbp}, however for the chunk starting at GPS time $1262192988 \, \SI{}{s}$ we observe a drop in efficiency of unknown origin.

\begin{figure}[t]
	\centering
	\includegraphics[scale = 1.]{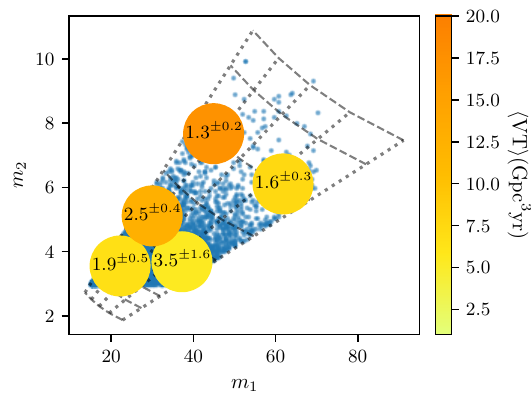}
	\caption{Search sensitivity computed over different regions of the parameter space. Each dot marks the center of a set of log-normally distributed injections and it is colored according to the sensitive space-time volume measured using such injection set. The number over-impressed denotes the ratio between the sensitive space-time volume of our search and the one of an aligned spin search targeting the same space.}
	\label{fig:sensitivity_bullet_points}
\end{figure}

The efficiency is used to compute the averaged sensitive space-time volume $\langle VT \rangle$ of the search, defined as:
\begin{equation}
	\langle VT \rangle(\text{FAR}) = T \int_0^\infty \text{d}z \, \epsilon(\text{SNR}(z), \text{FAR}) \frac{\text{d}V}{\text{d}z} \frac{1}{1+z}
\end{equation}
where $T$ is the total observation time considered by the search and $\frac{\text{d}V}{\text{d}z} \frac{1}{1+z}$ is the co-moving volume element at a given redshift $z$.
In Fig.~\ref{fig:sensitivity_bullet_points}, we report the VT as a function of the parameter space of our search. The efficiency is averaged over the four chunks for which it was computed and it is assumed constant for all the observation time.
We also report the $\langle VT \rangle$ improvement brought by our search as compared to an aligned-spin search. For each centroid, we employ $\sim 3\times10^4$ injections, log-normally distributed around the center with a variance of $0.1$. All the sensitive volumes are estimated at a fiducial ${\text{FAR} = 1 /\text{yr}}$.

\begin{table}[t]

\begin{tabular}{ll}
\toprule
$\text{FAR} = 1/\textrm{yr}\;$ & $\phantom{a}$ \\
$\phantom{a}$ & $\; \langle VT \rangle_\textrm{O3a} = 4.10^{+0.25}_{-0.26}\;\; \mathrm{Gpc^{3}yr}$ \\
$\phantom{a}$ & $\; \langle VT \rangle_\textrm{O3b} = 3.2^{+1.5}_{-1.5}\;\; \mathrm{Gpc^{3}yr}$ \\
\midrule
\hline
$\text{FAR} = 2.36/\textrm{yr}\;$ & $\phantom{a}$ \\
$\phantom{a}$ & $\; \langle VT \rangle = 8.4^{+1.5}_{-4.6}\;\; \mathrm{Gpc^{3}yr}$ \\
\bottomrule
\end{tabular}

\caption{Sensitive space-time volume estimated for the BGP population at a nominal FAR of $1/\text{year}$ and the FAR corresponding to the loudest event not reported in the GWTC. In the first case, we report the $\langle VT \rangle$ separately for O3a and O3b. }
\label{tab:VTs}
\end{table}

\subsection{Rate upper limits}\label{sec:rates}

The lack of newly detected signals allows to place an upper limit~\cite{Biswas:2007ni, PhysRevD.91.023005, LIGOScientific:2016kwr, LIGOScientific:2016ebi} on the presence of a subpopulation of asymmetric, heavily precessing signals, not identified by other searches.
%
In our analysis, we assume that such population is well described by the BGP injection set and, unlike standard works in BBH population analysis \cite{LIGOScientific:2020kqk,KAGRA:2021duu}, we do not attempt to constrain the distribution of BBH masses or spins.

After removing from consideration all the triggers associated to any GWTC event, the loudest trigger has a $\text{FAR} = 2.367 /\textrm{yr}$, as shown in Tab.~\ref{tab:significant_events}.
We can then use our knowledge of the sensitive space-time volume to place an upper limit $R_{90\%}$ (90\% confidence interval) on the merger rate of the said population of heavily precessing signals~\cite{Biswas:2007ni}:
\begin{equation}
	R_{90\%} = \frac{2.3}{\langle VT \rangle}
\end{equation}
where the sensitive space-time volume is estimate using the $\text{FAR}$ of the loudest event.
Using the VT values for the BGP population reported in Tab.~\ref{tab:VTs}, we obtain an upper limit of
\begin{equation*}
	R_{90\%} = 0.28^{+0.33}_{-0.04}\;\; \mathrm{Gpc^{-3}yr^{-1}}
\end{equation*}

The rate upper limit we obtain is consistent with the LVK rate estimates for our sub-population~\cite{KAGRA:2021duu}: considering that our target population is only a small fraction of the total BBH population considered in the analysis, the LVK results imply a BBH merger rate of
\begin{equation*}
	{R_\text{LVK} = 0.117-0.298 \; \text{Gpc}^{-3}\text{yr}^{-1}}
\end{equation*}
for our target region. Therefore our results do not provide any evidence to support the hypothesis of a population of asymmetric, heavily precessing signals that have not been detected by other searches.
Searching the data taken during the first and second observing runs~\cite{LIGOScientific:2019lzm} will increase the surveyed sensitive space-time volume, however it is unlikely that such increase will change the picture outlined.

According to our results, the past analysis carried on by the LVK have not missed any heavily precessing asymmetric system, despite having a poor sensitivity towards those signals. Therefore, the population analysis based on the confirmed events of the GWTC~\cite{LIGOScientific:2020kqk, KAGRA:2021duu, Hoy:2024qpy} retain their validity in light of our findings, at least in the mass region target by our search. In particular, we are able to confirm that precession is a rare event in BBHs, detectable in $\sim 2\%$ of the cases with the current detector's sensitivity, as pointed out in~\cite{Hoy:2024qpy}.



\section{Conclusion}

In this work, we presented the results of a matched-filtering search on the interferometric data from the LIGO-Virgo-KAGRA third observing run. Our search targets asymmetric precessing BBHs, with a doubled sensitivity over such extreme systems: the search method is thoroughly discussed in \cite{Schmidt:2024jbp}.
We report the detection of 30 candidates, already announced in the LIGO-Virgo-KAGRA (LVK) Transient Catalog, but we do not find any new candidate.
Our results allow us to place an upper limit on the merger rate of the binaries targeted by our search. As our upper limit is consistent with the rate estimates from the LVK, we can rule out the existence of a subpopulation of asymmetric precessing BBH not detected by past searches. Therefore, our results support the conclusions drawn from the existing catalogs: precessing systems are as ``rare'' as we thought they were before conducting the search.

Future work should focus on expanding the parameter space of the search, possibly targeting Neutron Star-Black Hole (NSBH) systems. 
This will serve as a valuable robustness test for our understanding of the spin properties in NSBH systems. The attempt will surely struggle with the huge size of the template bank required for this task, although see~\cite{McIsaac:2023ijd} which demonstrated that this is possible with only a 3x increase in template bank size.

        \begin{acknowledgments}
		S.S. is supported by the research program of the Netherlands Organisation for Scientific Research (NWO). S.C. is supported by the National Science Foundation under Grant No. PHY-2309332. The authors are grateful for computational resources provided by the LIGO Laboratory and supported by National Science Foundation Grants No. PHY-0757058 and No. PHY-0823459. This material is based upon work supported by NSF’s LIGO Laboratory which is a major facility fully funded by the National Science Foundation.
		LIGO was constructed by the California Institute of Technology and Massachusetts Institute of Technology with funding from the National Science Foundation (NSF) and operates under cooperative Agreement No. PHY-1764464. The authors are grateful for computational resources provided by the Pennsylvania State University’s Institute for Computational and Data Sciences (ICDS) and the University of Wisconsin Milwaukee Nemo and support by NSF Grant No. PHY-2011865, No. NSF OAC-2103662, No. NSF PHY-1626190, No. NSF PHY-1700765, No. NSF PHY-2207728, No. NSF PHY-2207594 and No. PHY-2309332 as well as by the research program of the Netherlands Organization for Scientific Research (NWO).
		This paper carries LIGO Document No. LIGO-P2400044. This research has made use of data or software obtained from the Gravitational Wave Open Science Center (gwosc.org), a service of LIGO Laboratory, the LIGO Scientific Collaboration, the Virgo Collaboration, and KAGRA. LIGO Laboratory and Advanced LIGO are funded by the United States National Science Foundation (NSF) as well as the Science and Technology Facilities Council (STFC) of the United Kingdom, the Max-Planck-Society (MPS), and the State of Niedersachsen/Germany for support of the construction of Advanced LIGO and construction and operation of the GEO600 detector. Additional support for Advanced LIGO was provided by the Australian Research Council. Virgo is funded, through the European Gravitational Observatory (EGO), by the French Centre National de Recherche Scientifique (CNRS), the Italian Istituto Nazionale di Fisica Nucleare (INFN) and the Dutch Nikhef, with contributions by institutions from Belgium, Germany, Greece, Hungary, Ireland, Japan, Monaco, Poland, Portugal, Spain. KAGRA is supported by Ministry of Education, Culture, Sports, Science and Technology (MEXT), Japan Society for the Promotion of Science (JSPS) in Japan; National Research Foundation (NRF) and Ministry of Science and ICT (MSIT) in Korea; Academia Sinica (AS) and National Science and Technology Council (NSTC) in Taiwan.
        \end{acknowledgments}

	\bibliography{biblio.bib}
	\bibliographystyle{ieeetr}

\end{document}